\documentclass[prb,twocolumn,showpacs,preprintnumbers,amsmath,amssymb]{revtex4}

\usepackage{graphicx}
\usepackage{dcolumn}
\usepackage{bm}

\begin{document}

\newcommand*{\cm}{cm$^{-1}$\,}
\newcommand*{\Tc}{T$_c$\,}


\title{Optical spectroscopy study on single crystalline LaFeAsO}
\author{Z. G. Chen}
\author{R. H. Yuan}
\author{T. Dong}
\author{N. L. Wang}

\affiliation{Beijing National Laboratory for Condensed Matter
Physics, Institute of Physics, Chinese Academy of Sciences,
Beijing 100190, China}


\begin{abstract}

Millimeter-sized single crystals of LaFeAsO were grown from NaAs
flux and the in-plane optical properties were studied over a wide
frequency range. A sizable electronic correlation effect was
indicated from the analysis of the free-carrier spectral weight.
With decreasing temperature from 300 K, we observed a continuous
suppression of the spectral weight near 0.6 eV. But a
spin-density-wave gap formation at lower energy scale was seen
only in the broken-symmetry state. We elaborate that both the
itinerancy and local spin interactions of Fe\emph{3d} electrons
are present for the FeAs-based systems; however, the establishment
of the long-range magnetic order at low temperature has a
dominantly itinerant origin.

\end{abstract}



\maketitle

The discovery of superconductivity in the
ferropnictides\cite{Kamihara08} has created tremendous interests
in the scientific community. While the initial research has mainly
concentrated on the compounds with ZrCuSiAs (1111)-type structure,
e.g., \textit{R}FeAsO$_{1-x}$F$_x$ (\emph{R}=La, Ce, Sm, Pr, Nd,
etc.), the focus in the community has been shifted to other
structural type compounds, e.g., AFe$_2$As$_2$ (122), Fe(Te,Se)
(11) and Na$_x$FeAs (111) systems. This is mainly due to the lack
of large-size single-crystal samples in 1111 systems. The crystal
growth in 1111 systems has been proven to be difficult. Despite
tremendous efforts over the past one year and a half, the single
crystals grown by various attempts or methods are limited to small
dimensions, typically less than 300 $\mu$m.\cite{Zhigadlo,HSLee}
As the superconductivity in Fe-pnictides was first discovered in a
1111 system\cite{Kamihara08} and the highest \emph{T}$_c$ was also
achieved in this structural type,\cite{Ren1} the 1111 systems are
of special interest. Very recently, Yan et al.\cite{JQYan}
reported a successful growth of millimeter-sized single crystals
of pure and doped LaFeAsO from NaAs flux, which opens the door to
explore the physical properties of single-crystal samples of 1111
systems.

Among different physical measurement techniques, optical
spectroscopy is powerful in probing the electronic excitations and
charge dynamics. It plays a key role in identifying an energy gap
and determining the spectral weight and transport lifetime of
quasi-particles. Optical studies on 1111 compounds have been done
by several groups since the discovery of superconducting
LaFeAsO$_{1-x}$F$_x$.\cite{DongEPL,Dubroka,DrechslerLa1,BorisLa,DrechslerLa2,Qazilbash}
Except for a work on LaFePO,\cite{Qazilbash} all measurements were
performed on polycrystalline samples. However, LaFePO appears very
different from LaFeAsO as no structural and magnetic instabilities
were found for it, while the measurements on polycrystalline
samples always mix the electronic contributions from the \emph{ab}
plane and the \emph{c} axis which could be very different for the
quasi-two-dimensional materials, thus leading to ambiguous
information. For example, the in-plane plasma frequency is a
quantity containing important information about carrier density or
many-body effect on the effective mass. As the overall reflectance
measured on polycrystalline sample is rather low and drops quickly
with increasing frequency, for example, \emph{R}($\omega)$ drops
to a value of $\sim$0.2 at about 3000 \cm, it is hard to obtain a
reliable value of plasma frequency from those data and, in fact,
the values estimated by different groups are rather
scattered.\cite{DrechslerLa1,BorisLa} In this work we report an
in-plane optical spectroscopy study on a parent LaFeAsO crystal
grown by NaAs flux. We show that the optical spectra of the
single-crystal sample are very different from the early
measurements on polycrystalline samples, but resemble closely to
the in-plane optical response in other FeAs-based parent systems
over broad frequencies. The study enables us to identify a strong
correlation effect and a partial energy gap formation in the
spin-density-wave (SDW) state for LaFeAsO.

The LaFeAsO single crystals used in this work were grown from a
procedure similar to that described by Yan et al.\cite{JQYan}
NaAs, which was found to be an effective solvent for
\textit{R}FeAsO, was prepared first by reacting Na and As chunks
at 300 $^o$C in a Al$_2$O$_3$ crucible sealed in quartz tube.
LaFeAsO from the mixture of LaAs, 1/3 Fe$_2$O$_3$, 1/3 Fe were
mixed with NaAs flux in a molar ratio of 1:20 and sealed in a Ta
tube under 1/2 atmosphere of argon gas. The Ta tube was then
sealed in an evacuated quartz tube and heated to 1180 $^o$C at a
rate of 60 $^o$C/h. After holding at this temperature for 20 h, it
was cooled to 600 $^o$C at 3$^o$C/h, then quenched to room
temperature. Plate-like crystals with surface dimensions of
(2-3)$\times$3 could be easily obtained by rinsing flux with
water.

Figure 1 shows the in-plane dc resistivity and magnetic
susceptibility as functions of temperature. The resistivity was
measured by a standard four-probe method. The magnetic
susceptibility was measured in a Quantum Design superconducting
quantum interference device vibrating-sample magnetometer under a
magnetic field of 1 T with \emph{H}$\parallel$ab-plane. The
resistivity decreases slightly with decreasing temperature from
300 to 250 K, then increases somewhat with further decreasing
temperature. A careful inspection indicates that the slope of the
\emph{T}-dependent resistivity $\rho(T)$ changes near 150 K then
shows a cusp near 132 K. $\rho(T)$ drops from 132 to 100 K then
increases fast at lower temperature. The two transitions near 150
and 132 K could be seen more clearly if we plot the derivative of
resistivity \emph{d}$\rho(T)$/\emph{d}\textit{T} as a function of
\emph{T}, shown in the inset of Fig. 1. They also manifest in
magnetic susceptibility as two weak steps. From the
neutron-scattering experiments,\cite{JQYan} it is clear that the
two transitions correspond to structural and magnetic transitions.
The overall resistivity and magnetic susceptibility are similar to
those seen in polycrystalline samples.\cite{DongEPL,GMZhang} The
stronger increase in resistivity at low temperature could be
ascribed to certain defects formed in the growth process.

\begin{figure}[t]
\includegraphics[width=8.5cm]{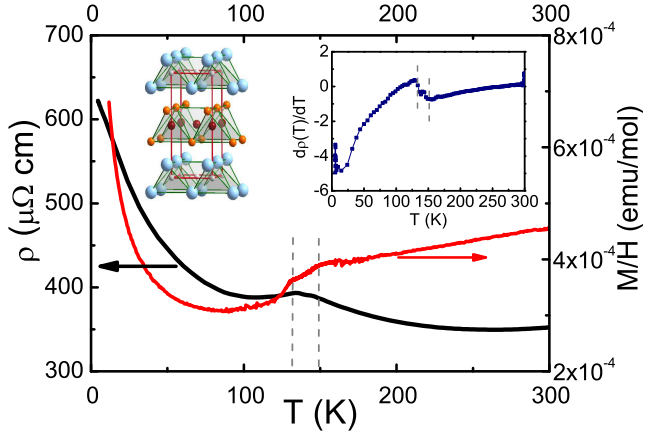}
\caption{(Color online) The dc resistivity and magnetic
susceptibility vs temperature for LaFeAsO single crystal. Two
anomalies could be identified at the dash lines positions. Inset:
the derivative of resistivity vs. temperature. A crystal structure
for LaFeAsO is also drawn.}
\end{figure}

The optical measurement was done on a Bruker IFS 66v/s
spectrometer on a fresh surface of 2$\times$2 mm$^2$ area in the
frequency range from 40 to 25000 cm$^{-1}$. An \textit{in situ}
gold and aluminum overcoating technique was used to get the
reflectance \emph{R}($\omega$). The real part of conductivity
$\sigma_1(\omega)$ is obtained by the Kramers-Kronig
transformation of \emph{R}($\omega$). Figure 2 shows the
room-temperature optical reflectance and conductivity spectra over
broad frequencies up to 25000 \cm. The overall spectral lineshapes
are very different from the data measured on polycrystalline
samples where the reflectance drops to a value below 0.2 near 3000
\cm,\cite{DrechslerLa1,DrechslerLa2} but here \emph{R}($\omega$)
is still higher than 0.6 near 3000 \cm. On the other hand, the
data on LaFeAsO crystal are very similar to those obtained on
AFe$_2$As$_2$ (A=Ba, Sr) single crystals.\cite{Hu122,DWu,Moon} The
reflectance drops almost linearly with frequency at low-$\omega$
region, then merges into the high values of a background
contributed mostly from the interband transitions from the
mid-infrared to visible regime.

The conductivity spectrum $\sigma_1(\omega$) displays a Drude-like
component at low frequencies. An important quantity that could be
extracted from those data is the in-plane plasma frequency
$\omega_p$. Within a single-band nearest neighbor tight binding
model, the kinetic energy of the electrons could be obtained from
the summarization of the spectral weight from the free
carriers,\cite{Baeriswyl,Qazilbash} which is proportional to the
square of the plasma frequency. The plasma frequency could also be
directly obtained from the density-function band-structure
calculations.\cite{DrechslerLa2,Boeri} Then, the ratio of the
experimental kinetic energy and the theoretical kinetic energy
from band-structure calculations, which is equal to the ratio of
the square of the experimental plasma frequency and the square of
band theory plasma frequency,
\textit{K}$_{exp}$/\textit{K}$_{band}$=$\omega_{p,exp}^2/\omega_{p,band}^2$,
provides a measure for the renormalization effect from the
electron correlations. \textit{K}$_{exp}$/\textit{K}$_{band}$ is
close to unity for a simple metal, but is reduced to zero for a
strongly correlated Mott insulator. For Fe pnictides, the
next-nearest-neighbor interaction and the multiorbitals play
roles, the above relation should be modified. However, in the
absence of a detailed theory, we still use the relation for a
rough measure.

\begin{figure}[t]
\includegraphics[width=6.5cm,clip]{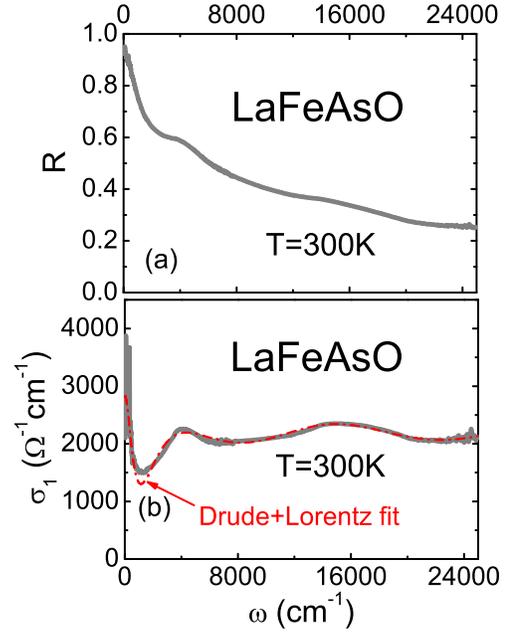}
\caption{(Color online) The room-temperature (a) optical
reflectance and (b) conductivity for LaFeAsO single crystal over
broad frequencies up to 25000 \cm. An analysis of the conductivity
spectrum by the Drude-Lorentz model is displayed as the dashed
curve. }
\end{figure}

From the partial sum rule, the effective carrier density per Fe
site below a certain energy $\omega$ can be obtained from
\begin{equation}
  \frac{m}{m^*}N_{eff}(\omega)=\frac{2mV_{cell}}{\pi{e^{2}}N}\int_{0}^{\omega}\sigma_1(\omega')d\omega',
\end{equation}
where \emph{m} is the free-electron mass, \emph{m}$^*$ is the
effective or renormalized optical mass, $V_{cell}$ is a unit-cell
volume, and \emph{N} is the number of Fe ions per unit volume.
$N_{eff}$ is related to an overall plasma frequency, after
choosing a proper high-frequency limit $\omega_c$, via the
relationship
$\omega_p^2=4\pi{e^{2}}N_{eff}(\omega_c)/m^*(V_{cell}/N)=8\int_{0}^{\omega_c}\sigma_1(\omega')d\omega'$.
Choosing $\omega_c\approx$1400 \cm, a frequency where
$\sigma_1(\omega)$ reaches its minimum but below the interband
transition, we get the plasma frequency
$\omega_p\approx$1.01$\times10^4$ \cm for \emph{T}=300 K.

An alternative way of making quantitative analysis is to fit the
conductivity spectrum by a Drude-Lorentz model to isolate the
different components of the electronic excitations,
\begin{equation}
\sigma_1(\omega)= {\omega_p^2 \over4\pi} {\Gamma_D\over\omega^2 +
\Gamma_D^2} + \sum{{S_j^2 \over4\pi}{\Gamma_j\omega^2 \over
(\omega_j^2-\omega^2)^2 + \omega^2\Gamma_j^2}}, \label{chik}
\end{equation}
where $\omega_p$ and $\Gamma_D$ are the plasma frequency and the
relaxation rate of conducting electrons, while $\omega_j$,
$\Gamma_j$, and $S_j$ are the resonance frequency, the damping,
and the mode strength of each Lorentz oscillator, respectively.
With a Drude component and three Lorentz oscillators centered at
4300, 16000 and 32000 \cm, describing the interband transitions,
the experimental data could be reasonably reproduced over a broad
frequency range, as seen in Fig. 2 (b). The fit yields
$\omega_p$=1.04$\times10^4$ \cm and $\Gamma_D$=640 \cm for
spectrum at 300 K. We find that the plasma frequency is in a rough
agreement with the value obtained by the sum-rule analysis. Note
that the plasma frequency for BaFe$_2$As$_2$ is around
1.3$\times10^4$ \cm in the nonmagnetic phase.\cite{Hu122} So
LaFeAsO has a slightly smaller plasma frequency. The
band-structural calculations give the \emph{ab} plane plasma
frequency of 2.1-2.3 eV.\cite{DrechslerLa2,Boeri} Then we can
estimate that
\textit{K}$_{exp}$/\textit{K}$_{band}$=$\omega_{p,exp}^2/\omega_{p,band}^2$$\approx$0.30-0.38.
This value is smaller than LaFePO,\cite{Qazilbash} indicating a
stronger reduction in the kinetic energy of the electrons with
respect to the band theory. This yields evidence that LaFeAsO is
more strongly correlated than LaFePO.

\begin{figure}[t]
\includegraphics[width=8.5cm,clip]{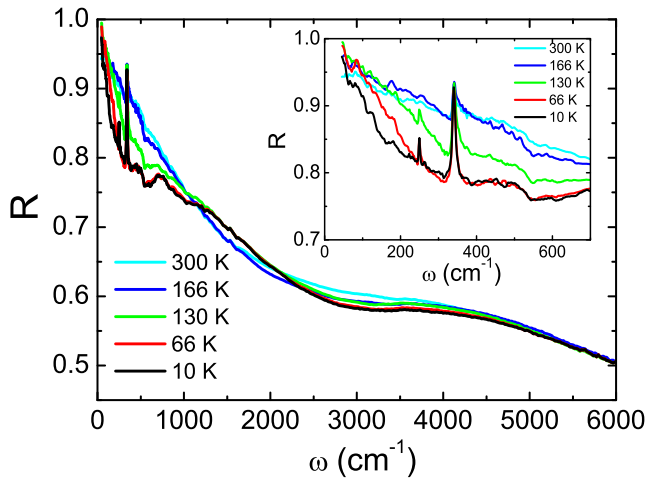}
\caption{(Color online) The \emph{ab} plane reflectance spectra
for LaFeAsO below 6000 \cm at several different temperatures.
Inset: the reflectance spectra in the low-frequency region.}
\end{figure}

We now turn to the temperature-dependent evolution of the optical
spectra. Figure 3 shows the reflectance R($\omega$) below 6000 \cm
at several different temperatures. The inset shows the
\emph{R}($\omega$) spectra in the range of 0-700 \cm. The spectra
start to show strong temperature dependence below 5000 \cm
($\sim$0.6 eV). Like BaFe$_2$As$_2$,\cite{Hu122}
\emph{R}($\omega$) is suppressed below 5000 \cm with decreasing
temperature, which also leads to the suppression in optical
conductivity spectra as shown in Fig. 4. At such high energies,
most of the spectral weight should come from the interband
transitions; however, the temperature-dependent part must have a
different origin. Because this suppression is seen well above the
magnetic and structural phase transitions and is present even at
room temperature and, furthermore, the energy of 0.6 eV is rather
high, it is not directly related to the establishment of the
long-range magnetic order. At present the origin of this
temperature-dependent structure remains unclear.

\begin{figure}[t]
\includegraphics[width=8.5cm,clip]{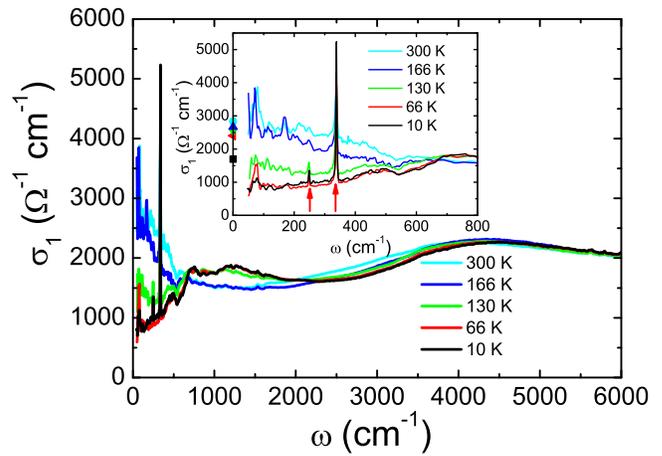}
\caption{(Color online) The \emph{ab} plane conductivity spectra
for LaFeAsO below 6000 \cm at several different temperatures.
Inset: an expanded plot of the conductivity spectra in the
low-frequency region. The two arrows indicate two phonon peaks.
The dc conductivity values at respective temperatures are also
shown.}
\end{figure}

More prominent spectral change is seen at lower frequencies when
the sample was cooling through the antiferromagnetic SDW
transition. The reflectance below 1000 \cm shows a remarkable
suppression, while it is enhanced between 1000 and 2000 \cm. The
spectral change in \emph{R}($\omega$) leads to the redistribution
of spectral weight in optical conductivity as seen in Fig. 4. The
spectral weight is severely suppressed at low frequencies and the
missed spectral weight is transferred to absorption peaks at high
energies. This gives optical evidence for the gap formation on the
Fermi surfaces below the SDW transition. As we addressed
earlier,\cite{Hu122,HuReview} the shape of the spectral weight
transfer is governed by the case I coherence factor for a SDW
order. Since the reflectance still increases fast toward unity at
lower frequencies, residual free carriers or Drude component is
still left in the magnetic ordered state. Therefore, the Fermi
surfaces are not fully gapped below \emph{T}$_N$. All those
spectral features are very similar to 122-type AFe$_2$As$_2$
(A=Ba, Sr)\cite{Hu122} and 111-type Na$_{1-\delta}$FeAs single
crystals\cite{Hu111} across the SDW transitions, indicating the
generic properties of undoped FeAs-based systems.

It should be noted that, although the spectral feature of LaFeAsO
resembles to other FeAs-based compounds, \emph{R}($\omega$) does
not increase toward unity as fast as in AFe$_2$As$_2$ (A=Ba, Sr)
below the magnetic or structural transition,\cite{Hu122} and this
leads to relatively broad residual Drude component in the SDW
state. In the inset of Fig. 4, we show the optical conductivity
spectra in the far-infrared region together with the respective dc
conductivity values. A good agreement between the dc and optical
data at the low-frequency limit could be directly seen at high
temperatures. In the low-temperature magnetic state, an upturn
feature of the Drude component should be present below our
measurement frequencies. However, comparing with the measurement
on BaFe$_2$As$_2$ crystal,\cite{Hu122} the increase in the
residual Drude component is less pronounced. The difference is
related to the different temperature-dependent behaviors of the dc
resistivity. BaFe$_2$As$_2$ crystal shows a steep decrease in
resistivity below the structural and magnetic transitions, whereas
the present LaFeAsO crystal shows only a slight decrease below the
magnetic transition; then the resistivity increases rapidly at
lower temperature. The peculiar property of the present sample is
attributed to the defects formed in the crystal growth.

The origin of the magnetism in the parent compounds of
Fe-pnictides is one of the focuses in the current research. The
above data offer further insight into this issue. The stripe or
collinear-type antiferromagnetic order was predicted correctly
from the nesting of disconnected electron and hole Fermi surfaces
being separated by a ($\pi$, $\pi$) wave vector at the very
beginning,\cite{DongEPL} and the importance of electron itinerancy
was also addressed by a number of theoretical works.
\cite{Mazin,Ran,Tesanovic,SinghReview} Alternatively, the
same-type magnetic structure could also be explained by a
local-spin Heisenberg exchange model by considering the nearest-
and next-nearest-neighbor superexchange
interactions.\cite{Yildirim,Si,Ma,FangXu,Wu} So the focus is
whether the parent compounds belong to the local or itinerant
category of antiferromagnets. Since the compounds are metals, the
itinerancy of electrons is apparently present. However, the strong
reduction in the kinetic energy with respect to the
density-function band-structural calculations clearly indicates
that the material is not a simple metal, but with strong electron
correlations. In general, we believe that both the itinerancy and
local-spin interactions of Fe\emph{3d} electrons are present.
Nevertheless, the long-range magnetic orders formed at low
temperature in FeAs-based compounds are dominantly driven by the
nesting tendency. This is because for all undoped FeAs-based
systems we observe clear SDW gap openings associated with the
transitions. This is expected for a nesting driven mechanism for
the broken symmetry state, but not the case for the local-spin
Heisenberg interactions. On the other hand, for 11-type parent
compound FeTe, the local-spin interaction physics seems to
dominate the magnetism as no SDW gap formation was observed below
the magnetic transition.\cite{Chen}

Finally, we briefly discuss the infrared phonon lines in the
far-infrared region. LaFeAsO crystallizes in the tetragonal
\emph{P}\textit{4/nmm} structure and according to the factor group
analysis, there are six infrared 3\emph{A}$_{2u}$ +3\emph{E}$_u$
active modes (along the \emph{c} axis and \emph{ab} plane,
respectively).\cite{Hadjiev} Experimentally we only observe two
modes at 248 and 338 \cm in the \emph{ab} plane response (see the
inset of Fig. 4). The latter mode is particularly strong in
intensity. Similar to the case of infrared measurement on the
polycrystalline samples,\cite{DongEPL} we do not observe any
splitting or new phonon lines below the structural or magnetic
transition. According to the \emph{ab initio} calculations on
LaFeAsO, the splitting of the infrared active phonon modes would
be very small and therefore could not be well resolved by the
optical measurement.\cite{Yildirim} On the other hand, the
intensities of the two modes show significant temperature
dependence. They are strongly enhanced below the structural or
magnetic transition. This behavior is similar to the observation
in BaFe$_2$As$_2$ crystal where two infrared active in-plane modes
near 94 and 253 \cm are observed. The two modes are expected for
the 122-structure type with \emph{I}\textit{4/mmm} space
group\cite{Litvinchuk} and the 253 \cm mode shows a strong
intensity enhancement below the structural or magnetic transition.
The intensity change was attributed to the charge redistributions
at each atom leading to a change in bonding between different
atoms.\cite{Akrap} Obviously, the 248 \cm mode in LaFeAsO is
associated with the Fe-As bonds, while the 340 \cm mode comes from
the vibration of oxygen.

To summarize, we studied the \emph{ab} plane optical properties of
LaFeAsO. We show that the optical spectra are very different from
the early measurements on polycrystalline samples, but resemble
closely to the in-plane optical response in other FeAs-based
parent systems over broad frequencies. Our analysis indicates a
reduction in kinetic energy of the electrons with respect to the
density-function band-structure calculations, thus evidencing a
relatively strong correlation effect. A temperature-dependent
spectral weight suppression is seen at rather high energy scale,
$\sim$0.6 eV, both below and above magnetic and structural
transitions. A further partial gap opening at lower energy scales,
with a significant spectral weight transfer from the free-carrier
region to energies above this gap, is seen only at low temperature
in the broken symmetry state. We elaborate that both the
itinerancy and local spin interactions of Fe\emph{3d} electrons
are present in Fe-based systems, but they have different dominant
contributions to the magnetic instabilities for different systems.
The long-range magnetic orders formed at low temperature in
FeAs-based compounds are dominantly driven by the nesting tendency
of disconnected Fermi surfaces.

\begin{acknowledgments}
We thank S. K. Su and Y. R. Zhou for their help in magnetization
measurement. This work was supported by the National Science
Foundation of China, the Knowledge Innovation Project of the
Chinese Academy of Sciences, and the 973 project of the Ministry
of Science and Technology of China.

\end{acknowledgments}


\end{document}